\documentclass[aps,twocolumn,amsmath,amssymb,superscriptaddress,floatfix,nofootinbib]{revtex4-1}
\usepackage{graphicx}
\usepackage{hyperref}
\usepackage{color}
\usepackage{lineno}              
\newcommand{\be}{\begin{equation}}
\newcommand{\ee}{\end{equation}}    
\newcommand{\beq}{\begin{eqnarray}}
\newcommand{\eeq}{\end{eqnarray}}
\newcommand{\ba}{\begin{array}}
\newcommand{\ea}{\end{array}}

\begin{document}

\title{Measurement of the helicity asymmetry  
$\mathbb{E}$ for the $\vec{\gamma}\vec{p} \to p \pi^0$ \\reaction in the resonance region}

\date{\today}

\newcommand*{\ANL}{Argonne National Laboratory, Argonne, Illinois 60439, United States of America}
\newcommand*{\ANLindex}{1}
\affiliation{\ANL}
\newcommand*{\ASU}{Arizona State University, Tempe, Arizona 85287, United States of America}
\newcommand*{\ASUindex}{2}
\affiliation{\ASU}
\newcommand*{\CSUDH}{California State University, Dominguez Hills, Carson, California 90747, United States of America}
\newcommand*{\CSUDHindex}{3}
\affiliation{\CSUDH}
\newcommand*{\CANISIUS}{Canisius College, Buffalo, New York 14208, United States of America}
\newcommand*{\CANISIUSindex}{4}
\affiliation{\CANISIUS}
\newcommand*{\CMU}{Carnegie Mellon University, Pittsburgh, Pennsylvania 15213, United States of America}
\newcommand*{\CMUindex}{5}
\affiliation{\CMU}
\newcommand*{\CUA}{The Catholic University of America, Washington, District of Columbia 20064, United States of America}
\newcommand*{\CUAindex}{6}
\affiliation{\CUA}
\newcommand*{\SACLAY}{IRFU, CEA, Universit\`{e} Paris-Saclay, F-91191 Gif-sur-Yvette, France}
\newcommand*{\SACLAYindex}{7}
\affiliation{\SACLAY}
\newcommand*{\CNU}{Christopher Newport University, Newport News, Virginia 23606, United States of America}
\newcommand*{\CNUindex}{8}
\affiliation{\CNU}
\newcommand*{\UCONN}{University of Connecticut, Storrs, Connecticut 06269, United States of America}
\newcommand*{\UCONNindex}{9}
\affiliation{\UCONN}
\newcommand*{\DUQUESNE}{Duquesne University, Pittsburgh, Pennsylvania 15282, United States of America }
\newcommand*{\DUQUESNEindex}{10}
\affiliation{\DUQUESNE}
\newcommand*{\FU}{Fairfield University, Fairfield, Connecticut 06824, United States of America}
\newcommand*{\FUindex}{11}
\affiliation{\FU}
\newcommand*{\FERRARAU}{Universit\`a di Ferrara, 44121 Ferrara, Italy}
\newcommand*{\FERRARAUindex}{12}
\affiliation{\FERRARAU}
\newcommand*{\FIU}{Florida International University, Miami, Florida 33199, United States of America}
\newcommand*{\FIUindex}{13}
\affiliation{\FIU}
\newcommand*{\FSU}{Florida State University, Tallahassee, Florida 32306, United States of America}
\newcommand*{\FSUindex}{14}
\affiliation{\FSU}
\newcommand*{\GWUI}{The George Washington University, Washington, District of Columbia 20052, United States of America}
\newcommand*{\GWUIindex}{15}
\affiliation{\GWUI}
\newcommand*{\GSIFFN}{GSI Helmholtzzentrum fur Schwerionenforschung GmbH, D-64291 Darmstadt, Germany}
\newcommand*{\GSIFFNindex}{16}
\affiliation{\GSIFFN}
\newcommand*{\INFNFE}{INFN, Sezione di Ferrara, 44100 Ferrara, Italy}
\newcommand*{\INFNFEindex}{17}
\affiliation{\INFNFE}
\newcommand*{\INFNGE}{INFN, Sezione di Genova, 16146 Genova, Italy}
\newcommand*{\INFNGEindex}{18}
\affiliation{\INFNGE}
\newcommand*{\INFNRO}{INFN, Sezione di Roma Tor Vergata, 00133 Rome, Italy}
\newcommand*{\INFNROindex}{19}
\affiliation{\INFNRO}
\newcommand*{\INFNTUR}{INFN, Sezione di Torino, 10125 Torino, Italy}
\newcommand*{\INFNTURindex}{20}
\affiliation{\INFNTUR}
\newcommand*{\INFNPAV}{INFN, Sezione di Pavia, 27100 Pavia, Italy}
\newcommand*{\INFNPAVindex}{21}
\affiliation{\INFNPAV}
\newcommand*{\IFKMA}{Institut f\"ur Kernphysik, Johannes Gutenberg University of Mainz, D-55099 Mainz, Germany}
\newcommand*{\IFKMAindex}{22}
\affiliation{\IFKMA}
\newcommand*{\KNU}{Kyungpook National University, Daegu 41566, South Korea}
\newcommand*{\KNUindex}{23}
\affiliation{\KNU}
\newcommand*{\LAMAR}{Lamar University, Beaumont, Texas 77710, United States of America}
\newcommand*{\LAMARindex}{24}
\affiliation{\LAMAR}
\newcommand*{\MIT}{Massachusetts Institute of Technology, Cambridge, Massachusetts  02139, United States of America}
\newcommand*{\MITindex}{25}
\affiliation{\MIT}
\newcommand*{\MISS}{Mississippi State University, Mississippi State, Mississippi 39762, United States of America}
\newcommand*{\MISSindex}{26}
\affiliation{\MISS}
\newcommand*{\ITEP}{National Research Centre Kurchatov Institute - ITEP, Moscow, 117259, Russia}
\newcommand*{\ITEPindex}{27}
\affiliation{\ITEP}
\newcommand*{\NMSU}{New Mexico State University, Las Cruces, New Mexico 88003, United States of America}
\newcommand*{\NMSUindex}{28}
\affiliation{\NMSU}
\newcommand*{\NSU}{Norfolk State University, Norfolk, Virginia 23504, United States of America}
\newcommand*{\NSUindex}{29}
\affiliation{\NSU}
\newcommand*{\NRC}{NRC Kurchatov Institute, PNPI, Gatchina 188300, Russia}
\newcommand*{\NRCindex}{30}
\affiliation{\NRC}
\newcommand*{\OHIOU}{Ohio University, Athens, Ohio 45701, United States of America}
\newcommand*{\OHIOUindex}{31}
\affiliation{\OHIOU}
\newcommand*{\ODU}{Old Dominion University, Norfolk, Virginia 23529, United States of America}
\newcommand*{\ODUindex}{32}
\affiliation{\ODU}
\newcommand*{\JLUGiessen}{II Physikalisches Institut, Justus Liebig University Giessen, 35392 Giessen, Germany}
\newcommand*{\JLUGiessenindex}{33}
\affiliation{\JLUGiessen}
\newcommand*{\MSU}{Skobeltsyn Institute of Nuclear Physics, Lomonosov Moscow State University, 119234 Moscow, Russia}
\newcommand*{\MSUindex}{34}
\affiliation{\MSU}
\newcommand*{\TEMPLE}{Temple University,  Philadelphia, Pennsylvania 19122, United States of America }
\newcommand*{\TEMPLEindex}{35}
\affiliation{\TEMPLE}
\newcommand*{\JLAB}{Thomas Jefferson National Accelerator Facility, Newport News, Virginia 23606, United States of America}
\newcommand*{\JLABindex}{36}
\affiliation{\JLAB}

\newcommand*{\ORSAY}{Universit\`{e} Paris-Saclay, CNRS/IN2P3, IJCLab, 91405 Orsay, France}
\newcommand*{\ORSAYindex}{37}
\affiliation{\ORSAY}
\newcommand*{\UNH}{University of New Hampshire, Durham, New Hampshire 03824, United States of America}
\newcommand*{\UNHindex}{38}
\affiliation{\UNH}
\newcommand*{\URICH}{University of Richmond, Richmond, Virginia 23173, United States of America}
\newcommand*{\URICHindex}{39}
\affiliation{\URICH}
\newcommand*{\ROMAII}{Universit\`a di Roma Tor Vergata, 00133 Rome Italy}
\newcommand*{\ROMAIIindex}{40}
\affiliation{\ROMAII}
\newcommand*{\SCAROLINA}{University of South Carolina, Columbia, South Carolina 29208, United States of America}
\newcommand*{\SCAROLINAindex}{41}
\affiliation{\SCAROLINA}
\newcommand*{\UTFSM}{Universidad T\`{e}cnica Federico Santa Mar\'{i}a, Casilla 110-V Valpara\`{i}so, Chile}
\newcommand*{\UTFSMindex}{42}
\affiliation{\UTFSM}
\newcommand*{\BRESCIA}{Universit\`{a} degli Studi di Brescia, 25123 Brescia, Italy}
\newcommand*{\BRESCIAindex}{43}
\affiliation{\BRESCIA}
\newcommand*{\UCR}{University of California Riverside, Riverside, California 92521, United States of America}
\newcommand*{\UCRindex}{44}
\affiliation{\UCR}
\newcommand*{\GLASGOW}{University of Glasgow, Glasgow G12 8QQ, United Kingdom}
\newcommand*{\GLASGOWindex}{45}
\affiliation{\GLASGOW}
\newcommand*{\YORK}{University of York, York YO10 5DD, United Kingdom}
\newcommand*{\YORKindex}{46}
\affiliation{\YORK}
\newcommand*{\VT}{Virginia Polytechnic Institute and State University, Blacksburg, Virginia 24061, United States of America}
\newcommand*{\VTindex}{47}
\affiliation{\VT}
\newcommand*{\VIRGINIA}{University of Virginia, Charlottesville, Virginia 22901, United States of America}
\newcommand*{\VIRGINIAindex}{48}
\affiliation{\VIRGINIA}
\newcommand*{\WM}{College of William and Mary, Williamsburg, Virginia 23187, United States of America}
\newcommand*{\WMindex}{49}
\affiliation{\WM}
\newcommand*{\YEREVAN}{Yerevan Physics Institute, 375036 Yerevan, Armenia}
\newcommand*{\YEREVANindex}{50}
\affiliation{\YEREVAN}
\newcommand*{\NOWJLAB}{Thomas Jefferson National Accelerator Facility, Newport News, Virginia 23606, United States of America}
\newcommand*{\NOWVIRGINIA}{University of Virginia, Charlottesville, Virginia 22901, United States of America}
\newcommand*{\NOWMERYLAND}{Montgomery College, Germantown, Maryland 20876, United States of America}

\author{\mbox{C.~W.~Kim}}
\affiliation{\GWUI}

\author{\mbox{N.~Zachariou}}
\altaffiliation{Corresponding author: \texttt{nick.zachariou@york.ac.uk}}
\affiliation{\YORK}

\author{\mbox{M.~Bashkanov}}
\affiliation{\YORK}

\author{\mbox{W.~J.~Briscoe}}
\affiliation{\GWUI}

\author{\mbox{S.~Fegan}}
 \affiliation{\YORK}

\author{\mbox{V.~L.~Kashevarov}}
\affiliation{\IFKMA}

\author{\mbox{K.~Nikonov}}
\affiliation{\NRC}

\author{\mbox{A.~Sarantsev}}
\affiliation{\NRC}

\author{\mbox{A.~Schmidt}}
\affiliation{\GWUI}

\author{\mbox{I.~I.~Strakovsky}}
\affiliation{\GWUI}

% ...

\collaboration{The CLAS Collaboration}
\noaffiliation

\begin{abstract}
The double-spin-polarization observable $\mathbb{E}$ for $\vec{\gamma}\vec{p}\to p\pi^0$ has been measured with the CEBAF Large Acceptance Spectrometer (CLAS) at photon beam energies $E_\gamma$ from 0.367 to $2.173~\mathrm{GeV}$ (corresponding to center-of-mass energies from 1.240 to $2.200~\mathrm{GeV}$) for pion center-of-mass angles, $\cos\theta_{\pi^0}^{c.m.}$, between -0.86 and 0.82. These new CLAS measurements cover a broader energy range and have smaller uncertainties compared to previous CBELSA data and provide an important independent check on systematics. These measurements are compared to predictions as well as new global fits from 
The George Washington University, Mainz, and Bonn-Gatchina groups. Their inclusion in multipole analyses will refine our understanding of the single-pion production contribution to the Gerasimov-Drell-Hearn sum rule and improve the determination of resonance properties. 
\end{abstract}

\maketitle

\section{Introduction}
\label{sec:intro}

The determination of resonance properties for all accessible baryon states is a central objective in nuclear physics. The extracted resonance parameters provide a crucial body of information for understanding the nucleon excitation spectrum and for testing models of the nucleon inspired by quantum chromodynamics (QCD) and, more recently, lattice QCD calculations. The spectra of $N^\ast$ and $\Delta^\ast$ baryon resonances have been extensively studied through meson-nucleon scattering and meson photoproduction experiments. Properties of the known resonances continue to become better determined as experiments involving polarized beams, targets, and recoil measurements are expanded and refined~\cite{CLAS:2017kua, Anisovich:2016vzt}. Extracted quantities include resonance masses, widths, branching fractions, pole positions, and associated residues, as well as photon decay amplitudes~\cite{ParticleDataGroup:2022pth}. The helicity 1/2 and 3/2 photon decay amplitudes
($N^* \to p\gamma$) can be extracted from a combination of resonance contributions to meson-nucleon scattering and photoproduction analyses.

New states have also been found, mainly through multichannel analyses that are sensitive to states having a relatively weak coupling to the $\pi N$ decay channel~\cite{Ronchen:2015vfa, CBELSATAPS:2014wvh, Kamano:2013iva}. 
A comprehensive overview of the available data is presented in Ref.~\cite{Ireland:2019uwn}. For the reaction of interest, experimental data on the differential cross section, beam-spin asymmetry, recoil polarization, beam-target polarization observables and others have been established for a wide range in energies and angles~\cite{Ireland:2019uwn}. Data on the double polarization observable $\mathbb{G}$ for the same reaction have also been recently published by the CLAS Collaboration~\cite{Zachariou:G}. This analysis builds from previously published work (see Ref.~\cite{Zachariou:G} and Ref.~\cite{CLAS:2015ykk}) using the same experiment and employing similar analysis procedures. Similar approaches were also employed by the CLAS Collaboration for the determination of the beam helicity asymmetry measurements using polarized neutrons~\cite{CLAS:2017kua, CLAS:2018gxz, CLAS:2020spy}.

Here, we have extracted the beam-target ($\mathbb E$) observable for neutral pion photoproduction from data taken with the CLAS FROzen Spin Target (FROST)~\cite{Keith:2012ad}. This observable is valuable both in multipole analysis and in providing a contribution to the Gerasimov-Drell-Hearn (GDH) and related sum rules~\cite{Strakovsky:2022tvu}.  Our extraction uses a different beam and polarized target apparatus as well as reaction identification methodology from the single previous measurement (see Ref.~\cite{CBELSATAPS:2013btn}). Where the data overlap with the previous measurement it provides an important independent check on systematics in the extraction of double-polarization observable $\mathbb{E}$, while improving the statistical quality of the world dataset. The new data also provide first information for energies $W<1.42~\mathrm{GeV}$.

As described below, the observable $\mathbb E$ is measured using a longitudinally polarized target and a circularly polarized photon beam. The difference of cross sections for helicity states $3/2$ and $1/2$, that is,  
$$\Delta(d\sigma/d\Omega) = (d\sigma_{3/2}/d\Omega - d\sigma_{1/2}/d\Omega) \>,$$ 
for $\vec{\gamma}\vec{N}\to N\pi $, is given in terms of helicity amplitudes:
\begin{equation}
	\frac{d\sigma_{3/2}}{d\Omega} = \frac{q}{k}\left(|H_1|^2 + |H_3|^2\right) \>, 
\end{equation}
\begin{equation}
	\frac{d\sigma_{1/2}}{d\Omega} = \frac{q}{k}\left(|H_2|^2 + |H_4|^2\right) \>,
	\label{eq:Hi}
\end{equation}
where $q$ and $k$ are the pion and photon center-of-mass momenta, respectively.
Helicity amplitudes are labeled following Ref.~\cite{Walker:1968xu} with $H_1$ and $H_3$ having
initial helicity 3/2 and final helicities 1/2 and -1/2, respectively; $H_2$ and $H_4$ have
initial helicity 1/2 and final helicities 1/2 and -1/2, respectively.
An integral involving $\Delta(d\sigma/d\Omega)$ gives the single-pion production part of the GDH sum rule~\cite{Strakovsky:2022tvu}. The sum and difference of the helicity $1/2$ and $3/2$ components can then be used to construct the beam-target polarization quantity $\mathbb E$~\cite{Barker:1975bp, Workman:1991xs} as
\begin{equation}
	\mathbb{E}~=~\frac{|H_2|^2 + |H_4|^2 - |H_1|^2 - |H_3|^2} {|H_2|^2 + |H_4|^2 + |H_1|^2 + |H_3|^2} \>.
    \label{eq:E}
\end{equation}

In order to extract the four helicity amplitudes, given that they are complex quantities, we need more than the cross section and $\mathbb E$. There are 16 possible experiments involving polarized beams, targets, and recoil particles, not all of which are independent~\cite{Barker:1975bp, Chiang:1996em}. However, the moduli of the helicity amplitudes can be determined with two additional double-polarization measurements (beam-recoil and target-recoil).
A complete solution for these amplitudes is phrased as the ``complete experiment'' problem, a topic that continues to be studied~\cite{Nakayama:2019cys}. The helicity amplitudes are constructed from an infinite sum of multipoles, and these are quantities that provide information on the existence and properties of resonances. This leads to the search for an appropriately truncated set of multipoles---a problem different from pursuing a complete 
experiment~\cite{Svarc:2017yuo, Wunderlich:2017dby, Workman:2016irf}. In this discussion, it should be remembered that these rules for finding multipoles are only guiding principles, as they ignore the influence of experimental uncertainties. In practice, all new experiments that improve the quality of existing measurements, or add information from new sources, are important to the program of multipole and resonance analysis. 
The advantage of the new CLAS FROST data presented here relative to previous CBELSA measurements~\cite{CBELSATAPS:2013btn} is the extended energy range covering lower energies with smaller uncertainties.


The paper is organized in the following manner. We give a brief background of the experimental conditions for this study in Sec.~\ref{sec:exp}. An overview of the method used to extract the double-polarized asymmetry results is given in Sec.~\ref{sec:reconstruction} and the uncertainty estimates for the data obtained are given in Sec.~\ref{sec:errs}. The resulting data are summarized and compared to various predictions in Sec.~\ref{sec:Result} and a new partial wave analysis (PWA), where we  compare multipoles obtained with and without including the present dataset is presented in Sec.~\ref{sec:PWA}. A summary and outlook are presented in Sec.~\ref{sec:sum}.

\section{Experiment}
\label{sec:exp}

The CLAS E--03--105 experiment~\cite{9a} (FROST or g9 run period) ran from December 2007 to February 2008 using the Continuous Electron Beam Accelerator Facility (CEBAF)~\cite{Leemann:2001dg} at Jefferson Lab in Newport News, Virginia. Data were collected using the CEBAF Large Acceptance Spectrometer (CLAS)~\cite{CLAS:2003umf} housed in Hall~B. This magnetic spectrometer allowed the efficient reconstruction of charged particles with polar angles between 8$^\circ$ and 140$^\circ$ over a large azimuthal acceptance ($\sim 83\%$). The spectrometer was constructed around a toroidal magnetic field and was comprised of drift chambers~\cite{Mestayer:2000we} for charged particle momentum determination, time-of-flight scintillators~\cite{Smith:1999ii} for particle identification, electromagnetic calorimeters~\cite{Amarian:2001zs} for neutral particle reconstruction, and a start counter~\cite{Sharabian:2005kq} that allowed the event start-time determination in photoproduction experiments.  Hall~B also housed the tagger spectrometer~\cite{Sober:2000we} that allowed the identification of the photon that initiated the reaction detected in CLAS, with energy resolution of $\Delta E\sim0.2\%$.

In this experiment, a circularly polarized tagged bremsstrahlung photon beam was incident on a longitudinally polarized proton target~\cite{Keith:2012ad} located near the center of the CLAS detector~\cite{CLAS:2003umf}.
The CEBAF electron beam was supplied at two different energies, 1.645 and $2.478~\mathrm{GeV}$. The electrons were delivered at currents between 33 and $45~\mathrm{nA}$ in beam bunches separated by about $2~\mathrm{ns}$. The electron beam helicity (and thus the photon helicity) was flipped at a rate of 
$30~\mathrm{Hz}$. The electron beam was incident on a $10^{-4}$ radiation-length thick gold foil radiator to produce the bremsstrahlung photon beam. The dipole magnet of the Hall~B photon tagger deflected the electron beam and post-bremsstrahlung electrons in order to tag photons produced with energies between $\sim$20\% and $\sim$95\% of the incident electron beam energy~\cite{Sober:2000we}. The degree of photon polarization varied between 20\% and 85\% depending on the incident electron beam energy and the bremsstrahlung photon energy. This was determined on an event-by-event basis using the Olsen and Maximon formula~\cite{Olsen:1959zz} 
 \begin{equation}\label{Eq:MxOls} 
        \begin{aligned}
            P_\odot & = P_e~\frac{4x - x^2}{4 - 4x + 3x^2} \>,
        \end{aligned}
    \end{equation}
where $x$ is the ratio of photon to electron energy $x = \frac{E_{\gamma}}{E_e}$ and $P_e$ is the electron polarization. The electron polarization was measured throughout the run period using the Hall~B M\o{}ller polarimeter~\cite{Moller}, and the average was established to be $P_e=0.835\pm0.035$.

The experiment utilized a FROzen Spin Target (FROST), made up of frozen butanol beads (C$_4$H$_9$OH), in which the protons in the hydrogen atoms were dynamically polarized. Butanol's covalently bonded protons in hydrogen atoms are polarizable using a technique called Dynamic Nuclear Polarization (DNP), in which spin polarization is simply transferred from the electrons to the nucleons.  Details of the target polarization procedure can be found in Ref.~\cite{Keith:2012ad}.

The dynamically polarized target resulted in polarization of the free protons within the butanol of over 90\%, with the polarization degrading over time -- typically about 1\% per day. Because of this, the target was re-polarized periodically. The degree of polarization of the free protons was determined on a run-by-run basis using Nuclear Magnetic Resonance (NMR) measurements~\cite{Goertz:2002vv}. The orientation of the spin of the free protons in the butanol target was also flipped regularly, enabling systematic checks.

Additional targets, carbon and polyethylene (CH$_2$), were placed downstream of the butanol target, which allows a detailed study of contributions from bound and unpolarized nucleons to our reaction yields. In practice, however, a free-proton signal was evident from the carbon target region, which was produced from hydrogen contamination (ice built up downstream of the target), and complicated this approach significantly. In this work, like in other FROST analyses~\cite{CLAS:2015ykk, CLAS:2015pjm, CLAS:2017yjv, CLAS:2017jrx, CLAS:2018mmb, CLAS:2021udy}, we report a result based on an analysis of the butanol target data alone. The secondary targets were only used to establish the systematic uncertainties related to contributions from unpolarized bound nucleons within the butanol target, as discussed in Sec.~\ref{Sec:BoundProt}.

\subsection{Double Polarization Observable $\mathbb{E}$}

This analysis is focused on the determination of the $\mathbb{E}$ observable, which manifests itself in the differential cross section in polarized beam - target experiments. In general, the differential cross section of polarized beam-target experiments for meson photoproduction reactions is given by~\cite{Ireland:2019uwn}
\begin{eqnarray*}
    \frac{d\sigma}{d\Omega} (E_\gamma, \cos\theta_{\pi^0}^{c.m.}, \phi) = \sigma_0[1-P_L\Sigma\cos(2\phi) \> \nonumber\\
    + P_x(-P_L\mathbb{H}\sin(2\phi) + P_\odot \mathbb{F}) \> \nonumber\\
    - P_y(-\mathbb{T} + P_L \mathbb{P} \cos(2\phi)) \> \nonumber\\
    - P_z(-P_L\mathbb{G}\sin(2\phi)+P_\odot \mathbb{E})] \>,
\end{eqnarray*}
where $P_L$ and $P_\odot$ correspond to the photon's degree of linear and circular polarization, and $P_x$, $P_y$, and $P_z$ correspond to the degree of target polarization along the $x$, $y$, and $z$ axes, respectively. Here, the $z$ axis points along the photon direction, and the $y$ axis is along the vertical direction in the lab frame. The azimuthal angle $\phi$ corresponds to the angle between the photon polarization vector (when the photon beam is linearly polarized) and the reaction plane defined by the incident photon and the outgoing pion directions. The observables $\Sigma$, $\mathbb G$, $\mathbb H$, $\mathbb T$, $\mathbb F$, $\mathbb P$, and $\mathbb E$ all depend on the kinematic variables $E_\gamma$ (the laboratory frame photon energy) and $\cos\theta_{\pi^0}^{c.m.}$ (the center-of-mass, $c.m.$, polar angle of the meson in the final state). For a circularly polarized beam ($P_L=0$ and $P_\odot\ne 0$) and target polarized along the $z$ direction ($P_x=0$ and $P_y=0$), the cross section equation reduces to   
\begin{eqnarray}
    \frac{d\sigma}{d\Omega} (E_\gamma, \cos\theta_{\pi^0}^{c.m.}) = \sigma_0[1-P_zP_\odot \mathbb{E}] \>.
\end{eqnarray}
Therefore, the observable $\mathbb E$ can be extracted from the unpolarized differential cross section $\sigma_0$ and the values of target and circular photon polarization $P_z$ and $P_\odot$, respectively. Alternatively, the observable $\mathbb E$ can be extracted from asymmetries utilizing various orientations of the target-photon polarization. Collecting data with both photon helicities and target polarizations along the $+z$ and $-z$ directions allows the cancellation of the detector acceptance and efficiency needed for the determination of the unpolarized cross section. Denoting the total helicity state (photon-target) with $1/2$ for the case where the photon helicity is anti-parallel (also denoted as $\uparrow\downarrow$) to the target polarization and $3/2$ for the case where the photon helicity is parallel to the target polarization (also denoted as $\uparrow\uparrow$~\footnote{The notations $\uparrow\downarrow$ and $\uparrow\uparrow$ represent the orientation of the target polarization relative to the photon helicity, with $\uparrow\downarrow$ and $\downarrow\uparrow$ being equivalent (same for $\uparrow\uparrow$ and $\downarrow\downarrow$ ). }), one can determine the observable $\mathbb E$ from
\begin{eqnarray}
   \mathbb{E} = \frac{1}{P_zP_\odot}\frac{\sigma^{1/2} - \sigma^{3/2}}{\sigma^{1/2} + \sigma^{3/2}} \>,
\end{eqnarray}
where $\sigma$ denotes the cross section of events obtained with the corresponding photon-target helicity. Assuming the detector efficiency, acceptance, and luminosity are constant throughout the experiment~\footnote{The effective acceptance for each configuration ({\it i.e.} $\uparrow\downarrow$ and $\uparrow\uparrow$) is the same due to the high beam-helicity flip rate of $30~\mathrm{Hz}$.}, the observable $\mathbb E$ can be directly extracted from the event yields ($N$) for each photon-target helicity configuration (the cross section is directly proportional to the event yield):
\begin{eqnarray}
   \mathbb{E} = \frac{1}{P_zP_\odot}\frac{N^{\uparrow\downarrow} - N^{\uparrow\uparrow}}{N^{\uparrow\downarrow} + N^{\uparrow\uparrow}} \>,
   \label{Eq:ExtractE}
\end{eqnarray}
where the detector and experimental effects listed above cancel out. It is evident from Eq.~(\ref{Eq:ExtractE}) that a detailed determination of the target polarization $P_z$ and photon polarization $P_\odot$ is needed for the precise determination of $\mathbb E$ (see Eq.~(\ref{Eq:MxOls})). 
It is worth noting here that Eq.~(\ref{Eq:ExtractE}) is valid when no contribution from unpolarized nucleons or background is present. We discuss the effect such contributions have on the determined observable and the need to determine a dilution factor in Sec.~\ref{Sec:BoundProt}.

\section{Reaction Reconstruction}
\label{sec:reconstruction}

Events with one positively charged track were retained for further analysis. We applied a set of selection cuts to the data to identify the $\vec{\gamma} \vec{p} \to p\pi^0$ reaction. The identification of final state protons from the sample of positively charged particles was performed by comparing a particle's speed, determined from time-of-flight ($\beta_m = \frac{distance}{time}$) and start counter information, to the particle's momentum, as determined from track curvature in the toroidal magnetic field. For a given momentum, $p$, the expected proton speed is given by $\beta_c = \frac{p}{\sqrt{p^2 + m^2}}$, where $m$ is the proton mass. Charged particles with $\Delta\beta = \beta_m - \beta_c$ around zero correspond to protons.

The CLAS reconstruction algorithms also allow the determination of the reaction vertex by extrapolating the particle's reconstructed track to the target region and evaluating the distance of closest approach with the incident beam position. The determined distance and time between the reaction vertex and the hit on the time-of-flight system allowed us to determine the vertex time of the event. Timing information from the tagger hodoscope also allowed us to determine the timing of each bremsstrahlung photon at the reaction vertex. Comparison between these two times allowed the unambiguous determination of the photon that initiated the reaction detected in CLAS.

Proton four-vectors were corrected for the expected energy loss sustained while exiting the target cell as well as for misalignments in the drift chambers and inaccuracies in the magnetic field maps (the latter corrections were established using the fully constrained reaction $\vec{\gamma} \vec{p} \to p\pi^+\pi^-$).

With this information, the reaction $\vec{\gamma} \vec{p} \to p\pi^0$ was fully reconstructed using the missing-mass technique. Figure~\ref{Fig:ExampleFits} shows the square of the missing mass of $\gamma p \to pX$ (labeled as $M^2_{\gamma p \to p X}$) for four kinematic bins. The clear peak around the squared mass of the neutral pion corresponds to the events of interest (photoproduction of $\pi^0$ off polarized protons). This peak sits on top of a smooth background. This is primarily caused by contributions from the photoproduction of $\pi^0$ off unpolarized and bound protons, which results in a wider missing-mass distribution due to the Fermi motion of the bound nucleon. Background from double pion photoproduction reactions was determined to have only a small contribution (1-3\%) using independent studies~\cite{CLAS:2013pcs}.

NMR measurements~\cite{Goertz:2002vv} allowed us to accurately determine the degree of proton polarization on a run-by-run basis. This reflects the polarization of events that originated from the free protons within the butanol target. The effective target polarization $P^{\mbox{\it{\tiny{eff}}}}_z$ allows us to account for events that originate from unpolarized material within the target cell. The determination of the effective target polarization was based on the relative yield between free- and bound-proton events. Contributions from bound nucleons dilute or reduce the effective target polarization, with the dilution factor, $D_F$, determined from the missing-mass distribution, as described below.

\subsection{Contributions from Bound Protons}\label{Sec:BoundProt}

The contributions from unpolarized bound protons within the target cell material (butanol) were accounted for in the analysis by the determination of the dilution factor. Considering the reaction $\vec{\gamma} \vec{p} \to p\pi^0$ originating from both free polarized and bound unpolarized protons, the yields obtained from these are given by:
\begin{eqnarray}
    N^{\uparrow\uparrow}_{free}   &=&N_0(1- P_z P_\odot \mathbb{E}) \>, \nonumber\\
    N^{\uparrow\downarrow}_{free} &=&N_0(1+ P_z P_\odot \mathbb{E}) \>, \nonumber\\
    N^{\uparrow\uparrow}_{bound}  &=&N'_0                           \>, \nonumber\\
    N^{\uparrow\downarrow}_{bound}&=&N'_0                           \>, \nonumber
\end{eqnarray}
where the experimental yield is given by 
$$N^{\uparrow\uparrow}_{exp}=N^{\uparrow\uparrow}_{free}+N^{\uparrow\uparrow}_{bound} \>,$$ and 
$$N^{\uparrow\downarrow}_{exp}=N^{\uparrow\downarrow}_{free}+N^{\uparrow\downarrow}_{bound} \>.$$ 
From this, the experimental asymmetry results in the following:
\begin{eqnarray}
    \frac{N^{\uparrow\downarrow}_{exp} - N^{\uparrow\uparrow}_{exp}}{N^{\uparrow\downarrow}_{exp}+N^{\uparrow\uparrow}_{exp} } = D_FP_zP_\odot \mathbb{E} \>, 
    \label{Eq:ObsE}
\end{eqnarray}
where $D_F = \frac{N_0}{N_0 + N'_0}$ is the dilution factor, and the product $D_FP_z$ is the effective target polarization $P^{\mbox{\it{\tiny{eff}}}}_z$.

In this analysis, the dilution factor was determined experimentally from the missing-mass distribution $\gamma p\to pX$, with $N_0$ representing the total yield of events from a free-proton target, and $N'_0$ the yield of events from bound protons. 
\begin{figure}[!ht]
\centering
        \includegraphics[width=8.8cm]{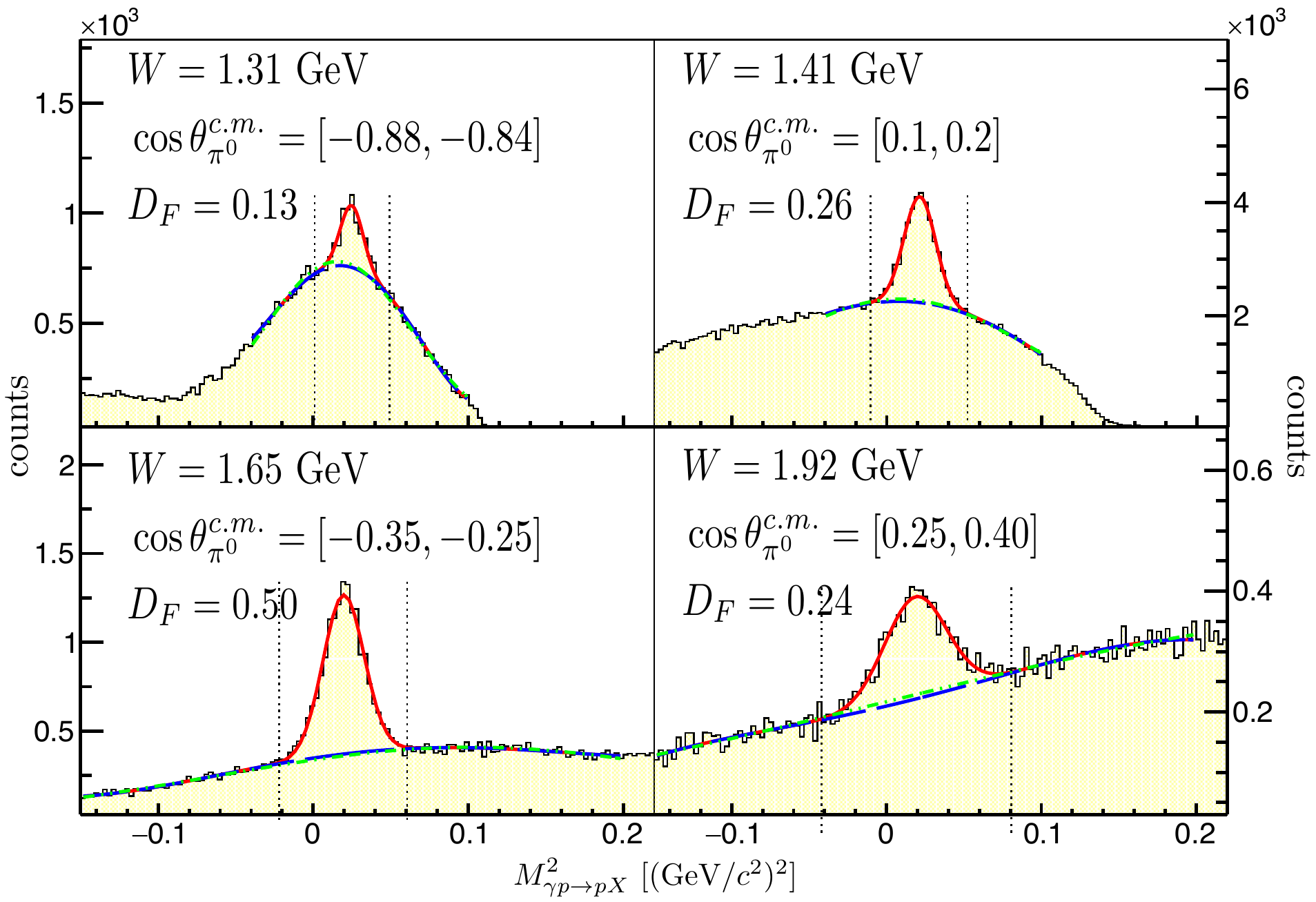}
        \caption{Missing mass squared distribution of $\gamma p\to pX$ for four kinematic bins. Two PDFs where used to determine the bound-nucleon contribution as shown with the blue dashed and green dash-dotted lines (red solid line shows the full fit). The vertical black dotted lines indicate the $\mu\pm3\sigma$ cuts applied to calculate the dilution factor and the reaction yields.}
        \label{Fig:ExampleFits}
\end{figure}
Specifically, we exploited the fact that reactions originating from bound protons results in a wider missing-mass distribution due to the Fermi motion of the target nucleon, to determine the ratio between free protons to our total yield. A carbon target upstream of the butanol target allowed us to establish the expected probability distribution function (PDF) that describes the missing mass from bound-nucleon events. The bound-nucleon PDF and a Gaussian to describe the free-proton events were fitted to the missing-mass distribution of events originating from the butanol target. The dilution factor was then determined by integrating the bound-nucleon PDF in the $\mu\pm3\sigma$ range established from the Gaussian fit (where $\mu$ and $\sigma$ are the mean and standard deviation of the free-proton peak). An example of such fits in four kinematic bins is shown in Fig.~\ref{Fig:ExampleFits}. Different bound-nucleon PDFs were utilized to systematically study the effect these have on the determination of the dilution factor, as described below.

\subsection{Yield Determination}

As mentioned before, the polarization observable $\mathbb{E}$ was determined using the asymmetry of yields from the two photon-target polarization configurations (parallel and anti-parallel), as shown in Eq.~(\ref{Eq:ObsE}). The yields correspond to the total number of events with a $z$-vertex cut between $-3~\mathrm{cm}$ and $3~\mathrm{cm}$ that enabled us to select events that originated within the butanol target and within a missing-mass range that was dependent on the kinematic bin. The missing-mass range was the same as the range used in the dilution factor determination, established from fits to the missing-mass with a Gaussian to describe the free-proton contributions and a bound-nucleon PDF (either a second Gaussian or a polynomial). The range was then established to be at $\mu\pm3\sigma$ for each kinematic bin.

\section{Uncertainties}
\label{sec:errs}

The statistical uncertainties of $\mathbb{E}$ were determined using  error propagation from the two yields, and accounting for the statistical uncertainty associated with the dilution factor determination. The latter was determined using the covariance matrix of the bound-nucleon fit parameters, as well as the integral and its uncertainty of the fit to the butanol missing-mass distribution.

\begin{table}[htbp]
\centering
\begin{tabular}{|c|c|}
\hline\hline 
Source & $\sigma^{sys}$\\
\hline\hline
Particle identification             & 0.002 \\
Reaction reconstruction / $m_X$ cut & 0.008 \\
Photon selection                    & 0.015 \\
Vertex cuts                         & 0.006 \\
Fiducial cuts                       & 0.002 \\
Dilution factor                     & 0.014 \\
Point-to-point $D_F$                & 0.0--0.3 \\
\hline
\bf{Total Point-by-point (absolute) Syst.} & \bf{0.023$-$0.301} \\
\hline
Photon polarization                 & 4\% \\
Target polarization                 & 6\% \\
\hline
\bf{Global Scale (relative) Syst.}        & \bf{7.2\%} \\
 \hline\hline
\end{tabular}
\caption{Summary of systematic uncertainties related to the determination of the double-polarization observable $\mathbb{E}$. 
\label{ref:TablSystUn} }
\end{table}
A thorough assessment of systematic effects in the determined observable was carried out, including effects related to particle identification and reaction reconstruction. Uncertainties in the photon and target polarization were also evaluated and included as a global scale systematic effect. Systematic uncertainties related to the dilution factor determination were also evaluated in detail. A total dilution factor systematic uncertainty of $\sim4\%$ was established by studying the effect different PDFs had on describing the bound-nucleon contributions. The dilution factor is expected to have a smooth dependence on the kinematic variables. A point-to-point dilution factor systematic uncertainty was also included to account for differences in the dilution factor from this smooth variation. This point-to-point systematic uncertainty was established using interpolation of the dilution values from adjacent bins and the determined value of the bin in question.

A summary of the systematic uncertainties is given in Table~\ref{ref:TablSystUn}. The uncertainties are split into a point-by-point (absolute) uncertainty that was applied to all points~\footnote{The point-by-point uncertainty was added to each point's statistical uncertainty in quadrature and it was treated independently for each point.}, and a relative uncertainty (associated with the target and photon polarizations) that was applied as a scale systematic affecting all points in a correlated way.

\section{Results and Discussion}
\label{sec:Result} 
\begin{figure}[!ht]
\centering
      \includegraphics[width=8.7cm]{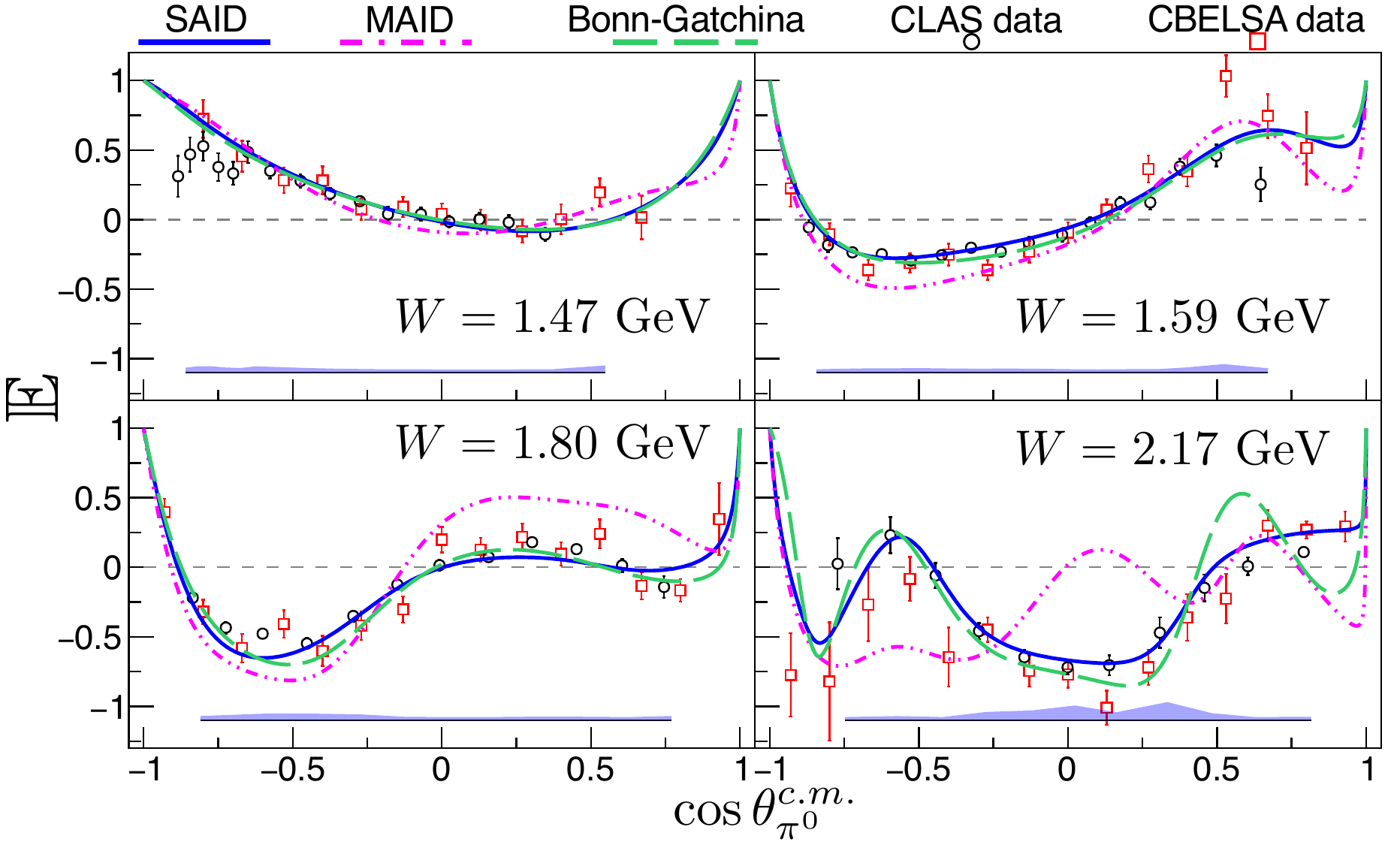}
\caption{\label{Fig:EPred} Results from CLAS (black open circles) for $\mathbb{E}$ as compared to published data from CBELSA~\cite{CBELSATAPS:2013btn} (red open squares) for four kinematic bins. Statistical and point-to-point systematic uncertainties are combined in quadrature. Systematic uncertainties for CLAS results are indicated by the shaded regions at the bottom of each plot.  PWA predictions for Scattering Analysis Interactive Database (SAID) SM22~\cite{Briscoe:2022wj}, MAID PIONMAID-2021~\cite{Kashevarov:2017vyl}, and Bonn-Gatchina BnGa-2022-02~\cite{CBELSATAPS:2022uad} are shown in solid blue, magenta dashed-dotted, and green dashed curves, respectively. The CBELSA data were included in the SAID, MAID, and 
Bonn-Gatchina fits.} 
\end{figure}

The current work enabled the determination of $\mathbb{E}$ between center-of-mass ($c.m.$) energies $W=1.25~\mathrm{GeV}$ and $W=2.23~\mathrm{GeV}$ for a wide angular coverage of the $\pi^0$. These results extend the kinematic reach of the world dataset for the  $\mathbb{E}$ observable to lower energies, while significantly improving the statistical precision in all energy bins. The newly obtained values agree well with previously published data from CBELSA~\cite{CBELSATAPS:2013btn}, as shown in Fig.~\ref{Fig:EPred}. Specifically, Fig.~\ref{Fig:EPred} shows the CLAS results on $\mathbb{E}$ (black open circles) for four $c.m.$ energy bins and how they compare to CBELSA (red open squares) results. The figure also provides predictions ({\it i.e.} PWA solutions where the newly obtained data were not included in the fits) from the SAID (solid blue line), the MAID (magenta dashed-dotted line), and the Bonn-Gatchina (green dashed line) PWA solutions. It is evident that at lower energies, where a plethora of other data exist~\cite{Ireland:2019uwn}, the PWAs predict well the precise measurement of $\mathbb{E}$ from CLAS. At higher energies, significant deviations between the solutions and data are evident. Overall, the SAID and Bonn-Gatchina PWA solutions agree well, especially at larger $\pi^0$ angles in the $c.m.$ frame, whereas deviations in all angles are observed between the MAID PWA solutions and our data. 

\begin{figure*}[!ht]
\includegraphics[width=18cm]{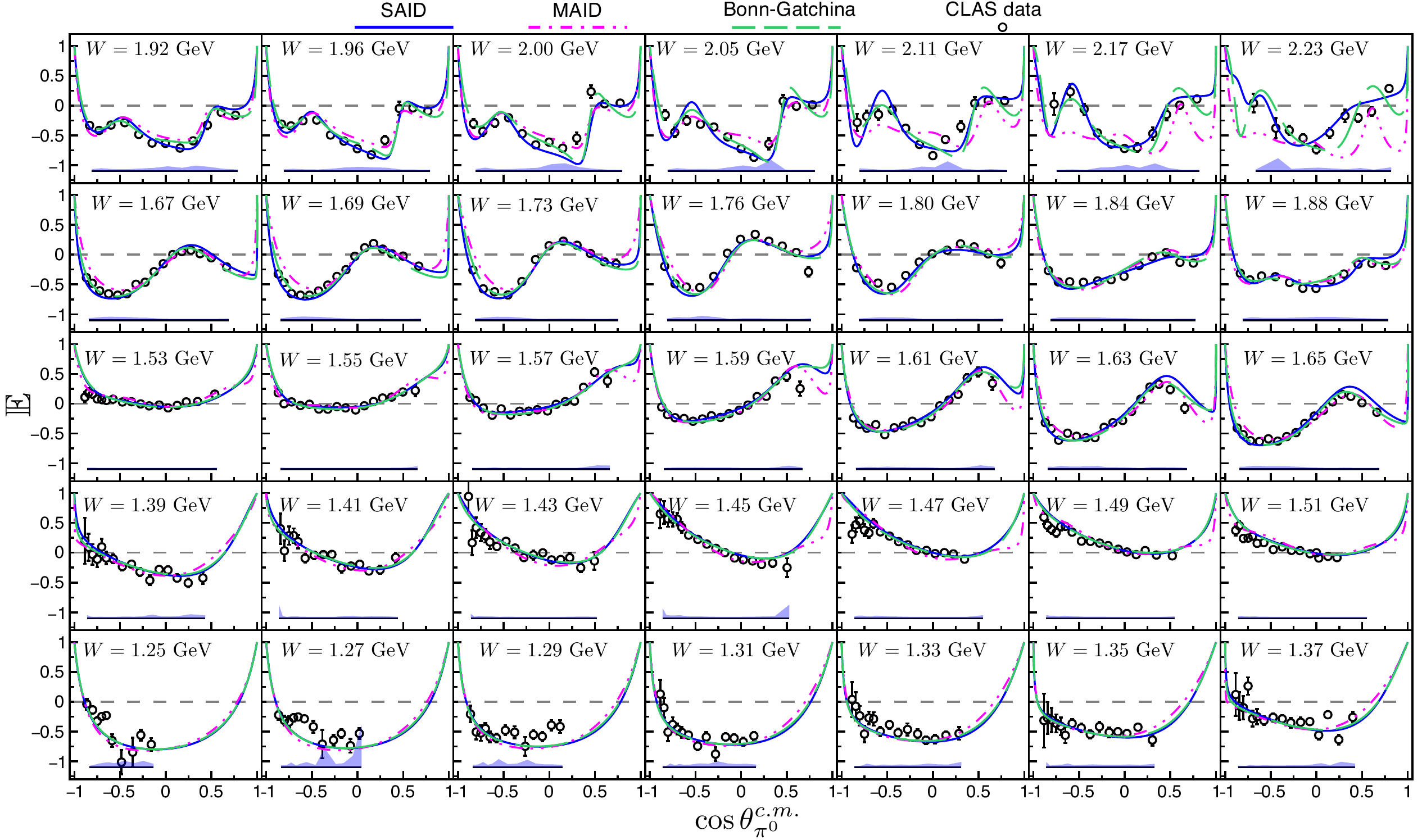}
\caption{\label{Fig:EFits} Double-polarization observable $\mathbb{E}$ (black open circles) from this work as a function of the pion angle in the $c.m.$ frame. The different panels denote bins in $c.m.$ energy $W$. The new SAID KI22 (blue solid curves), the Bonn-Gatchina (green dashed curves), and MAID (magenta dashed-dotted curves) solutions are also indicated in the panels. Statistical and point-to-point systematic uncertainties are combined in quadrature. Systematic uncertainties are indicated in the shaded regions at the bottom of each plot.}
\end{figure*}

The full kinematic coverage of the CLAS dataset is illustrated in Fig.~\ref{Fig:EFits} (CBELSA results are omitted in these plots). The new results were included in the world database and PWA fits were performed in all three frameworks. Specifically, the new PWA solutions for SAID (blue solid line), MAID (magenta dashed-dotted line), and Bonn-Gatchina (green dashed line) that account for the CLAS $\mathbb{E}$ results (in addition to the other world data), are shown in Fig.~\ref{Fig:EFits}. These new fits describe well the $\mathbb{E}$ data, with the exception of the MAID solution at energies $W>2.10~\mathrm{GeV}$. More details on the PWA fits and their findings are provided below. 

\subsection{Multipole analysis}
\label{sec:PWA} 
In the SAID multipole analysis of these data, an energy-dependent parametrization,
based on the Chew-Mandelstam $K$-matrix approach, has been used.
The Chew-Mandelstam parametrization (CM) for a hadronic $T$ matrix, described in Ref.~\cite{Arndt2006bf}, was used in a previous coupled-channel fit of $\pi N$ elastic scattering and $\pi N\to N\eta$ reaction data.  The parametrization form used in that fit was given as
\begin{equation}
\label{eqn:Tab}
    T_{\alpha\beta} = \sum_\sigma [1-\bar{K} C]^{-1}_{\alpha\sigma} \bar{K}_{\sigma\beta} \>,
\end{equation}
where the notation $\bar{K}$ was used to distinguish this from the Heitler $K$-matrix~\cite{Paris:2010tz}
and $\alpha$, $\beta$, and $\sigma$ are indices that label the channels, $\pi N$, $\pi \Delta$, $\rho N$, {\rm and} $\eta N$. The parameter $C$ corresponds to the Chew-Mandelstam function described in Ref.~\cite{Arndt:1985vj}.  Given the success of this approach in the hadronic two-body sector, the fit formalism was extended to pion photoproduction~\cite{Workman:2012hx}.

The Chew-Mandelstam form of Eq.~\eqref{eqn:Tab} has been extended to include the electromagnetic channel as:
\begin{equation}
\label{eqn:Tag}
    T_{\alpha\gamma} = \sum_\sigma [1-\bar{K} C]^{-1}_{\alpha\sigma} \bar{K}_{\sigma\gamma} \>.
\end{equation}
Here, $\gamma$ denotes the electromagnetic channel, $\gamma n$, and $\sigma$ denotes the hadronic channels that appear in the parametrization of the hadronic rescattering matrix,                        
    $$[1-\bar{K} C]^{-1} \>.$$   
By sharing this common factor, which qualitatively encodes the hadronic channel coupling (or rescattering) effects, Eqs.~\eqref{eqn:Tab} and \eqref{eqn:Tag} constitute a unified approach to the problem of parametrizing the hadronic scattering and photoproduction amplitudes.

The existing hadronic elements of Eq.~(\ref{eqn:Tab}) were not varied in the fits of the photoproduction data. For this reason, the photoproduction fits have a resonance structure identical to that found in Ref.~\cite{Arndt2006bf}. The electromagnetic CM $K$-matrix elements contain polynomials in energy with the correct threshold behavior. The order of these polynomials was increased until the fit's $\chi^2$ value was not significantly improved. At this level, increasing the order no longer produced significant improvements to the fit.
While the initial fit from SAID to the present set of data delivered a good overall description, some systematic deviations were noticed at back angles and at the highest energies. 


PionMAID-2021 is an updated version of the unitarity isobar model MAID2007~\cite{Drechsel:2007if}.  It has been developed to analyze the world data of neutral- and charged-pion photoproduction. The model contains a resonance part, parameterized by a Breit-Wigner shape, and a background with Born terms in the resonance region. Regge phenomenology was applied at energies above the resonance region~\cite{Kashevarov:2017vyl} for neutral and for charged pion photoproduction \cite{Guidal:1997hy, Laget:2019tou}.  The model describes experimental data up to photon energies of $18~\mathrm{GeV}$ and is well adapted for predictions at higher energies.

The Bonn-Gatchina model was developed for the partial wave analysis of reactions with multi-particle final states. The particle interaction vertices are described in the framework of the covariant tensor formalism and the energy-dependent part of the partial wave amplitudes satisfies unitarity and analyticity conditions. The description of the method can be found in Ref.~\cite{Anisovich:2011zz}.  The present solution describes 201 datasets, which includes the pion- and photon-induced reactions  with one or two pseudoscalar mesons in the final state, as well as photon-induced reactions with production of the $\omega$ meson.

\begin{figure}[!ht]
        \centering
        \includegraphics[width=0.47\textwidth,keepaspectratio]{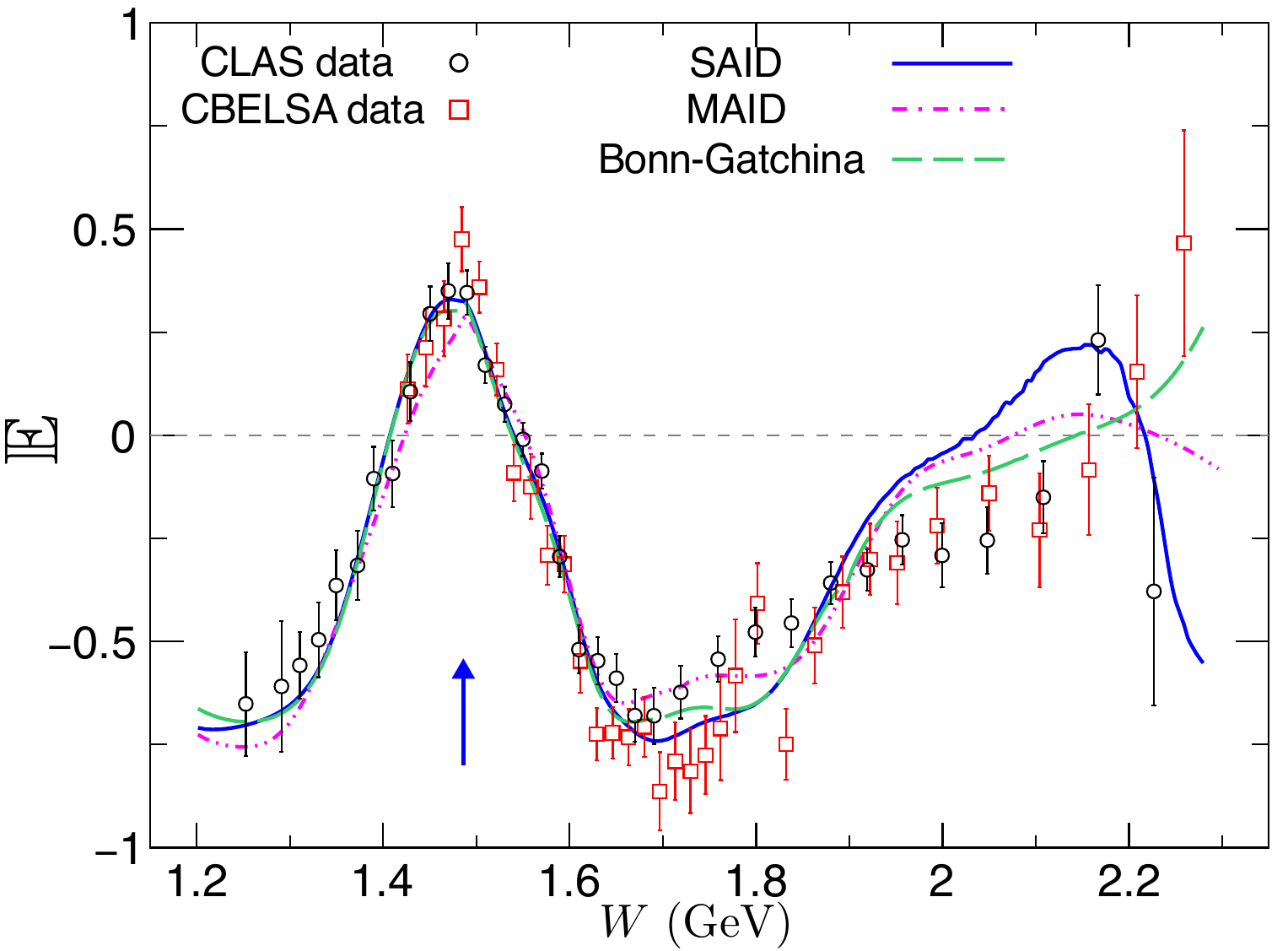}
        \caption{Comparison of the present CLAS FROST data (black open circles) and previous CBELSA measurements~\cite{CBELSATAPS:2013btn} (red open squares) vs. present SAID KI22 (blue solid curve), MAID (magenta dashed-dotted curve), and Bonn-Gatchina (green dashed curve) solutions for the $\vec{\gamma}\vec{p}\to p\pi^0$ reaction and for the double-polarization observable $\mathbb{E}$ at $120^\circ$. The new CLAS FROST $\mathbb{E}$ data are included in the fits. Statistical and point-to-point systematic uncertainties are combined in quadrature. The blue vertical arrow indicates the $\eta$ production threshold.}
        \label{fig:120}
    \end{figure}

\begin{figure}[!ht]
        \centering
        \includegraphics[width=0.47\textwidth,keepaspectratio]{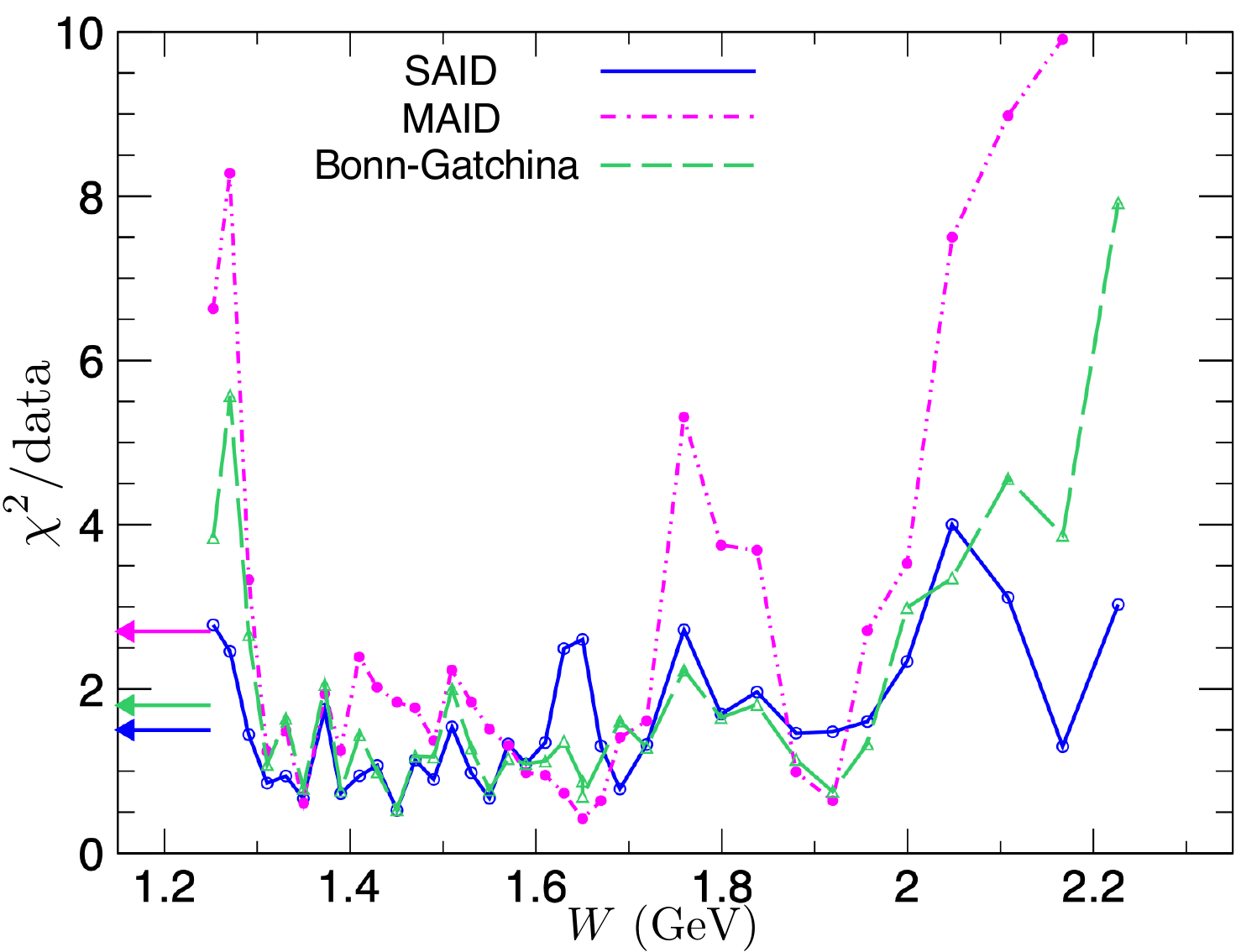}
        \caption{Comparison of the present SAID KI22 (blue open circles and solid line), MAID (magenta filled circles and dashed-dotted line), and 
        Bonn-Gatchina (green open triangles and dashed line) fits for the $\vec{\gamma}\vec{p}\to p\pi^0$ reaction for the double-polarization observable $\mathbb{E}$. The lines connecting the points are included only to guide the eye. Shown are the fit $\chi^2$-per-data-point values averaged within each energy bin, where the arrows (blue for SAID, magenta for MAID, and green for Bonn-Gatchina) show the overall $\chi^2$ per data point values from Table~\ref{ref:TablChi2}.}
        \label{fig:chi}
\end{figure}

In Table~\ref{ref:TablChi2}, $\chi^2$ values are compared for fits in which the present data were included and those in which they were not. Figure~\ref{Fig:EFits} gives a more detailed view of the improved quality of the fits obtained with the SAID, MAID and Bonn-Gatchina approaches. While the overall fit is quite good, the largest discrepancies are seen at backward angles and higher energies, as illustrated in Fig.~\ref{fig:120}.  Fig.~\ref{fig:chi} similarly shows that the fit is not uniformly good, with the largest $\chi^2$ values occurring at higher energies, as one would expect. 

As for variations in multipole amplitudes due to the inclusion of the new data, the Bonn-Gatchina group found changes in the helicity 1/2 and 3/2 photon decay amplitudes associated with the $N(2120)3/2^-$ and changes of the helicity 1/2 amplitude associated with the $\Delta (1940)3/2^-$. These states are not included in the SAID model and changes were seen only for amplitudes at the highest energies, which were small in magnitude. No significant resonance changes were reported in the MAID re-analysis.

\begin{table}[htbp!]
\centering
\begin{tabular}{|c|c|c|}
\hline\hline
PWA  & No FROST                    & With FROST \\
\hline\hline
SAID & SM22: \textbf{2.1}          & KI22: \textbf{1.5} \\
MAID & pionMAID-2021: \textbf{5.2} & \textbf{2.7} \\
BnGa & BnGa-2022-02:  \textbf{2.7} & \textbf{1.8} \\
 \hline\hline
\end{tabular}
\caption{Summary of $\mathrm{\chi^2/}$data point for the new FROST $\mathbb{E}$ data.  PWA solutions are 
SAID: SM22~\cite{Briscoe:2022wj} and new KI22.
MAID: PIONMAID-2021~\cite{Kashevarov:2022}.
Bonn-Gatchina: BnGa-2022-02~\cite{CBELSATAPS:2022uad}.
\label{ref:TablChi2} }
\end{table}

While the inclusion of new and precise polarization measurements has led to better agreement between groups extracting multipole amplitudes~\cite{Anisovich:2016vzt}, the prediction of new quantities, or existing quantities outside their ranges of measurement, is generally only qualitative, as can be seen in Figs.~\ref{Fig:EPred} and \ref{fig:120}. The most extensive single- and double-polarization data above photon energies of $2~\mathrm{GeV}$, come from pre-1980 Daresbury experiments with limited angular coverage. More recent measurements at Jefferson Lab, MAMI, Bonn, and SPring-8 have provided data with higher precision and broader angular range, but do not provide a database approaching a ``complete experiment.'' Each new measurement is particularly valuable at these higher energies. More detailed analyses from SAID~\cite{Briscoe:2022wj}, MAID, and Bonn-Gatchina are expected and will include these and other recent data.

\section{Summary and Outlook}
\label{sec:sum}

The CLAS E–03–105 experiment (FROST), which utilized longitudinally polarized protons and circularly polarized photons, allowed the precise determination of the beam-helicity asymmetry $\mathbb{E}$ for the $\vec{\gamma}\vec{p}\to p\pi^0$ reaction.  The new results provide an important independent check on the extraction of the double-polarization observable $\mathbb{E}$ with a precision improved over the past data from CBELSA, while extending the kinematical coverage at lower energies. Specifically, the newly obtained data cover $c.m.$ energies between $W=1.25~\mathrm{GeV}$ and $W=2.23~\mathrm{GeV}$, whereas results from CBELSA provided an energy coverage between $W=1.42~\mathrm{GeV}$ and 
$W=2.58~\mathrm{GeV}$.

PWA fits within the SAID, MAID, and Bonn-Gatchina frameworks were performed with the inclusion of this newly obtained dataset. The Bonn-Gatchina group found changes in the helicity 1/2 and 3/2 photon decay amplitudes associated with the $N(2120)3/2^-$ and changes of the helicity 1/2 amplitude associated with the $\Delta(1940)3/2^-$, while the MAID solution did not show any significant changes. A detailed analysis from the SAID group is underway to establish the impact of this new dataset and their findings will be reported in a dedicated paper.  


Numerical CLAS FROST $\mathbb E$ data are available in the SAID~\cite{SAID:DB}, CLAS~\cite{CLAS:DB}, and University of York Pure databases~\cite{Data-DOI}.

\vspace{5mm}
\section{Acknowledgments}

This work was supported in part by the U.~S.~Department of Energy, Office of Science, Office of Nuclear Physics under Awards No.~DE--SC0016583 and DE--SC0016582; by the U.~K.~Science and Technology Facilities Council grants, ST/P004385/2, ST/T002077/1, and ST/L00478X/2; by the EU Horizon 2020 research by innovation program, STRONG-–2020 project (under grant agreement No.~824093), as well as by the Russian Science Foundation RSF22--22--00722 grant. We also acknowledge the outstanding efforts of the staff of the Accelerator and Physics Divisions at Jefferson Lab that made this experiment possible. The Southeastern Universities Research Association (SURA) operated the Thomas Jefferson National Accelerator Facility for the United States Department of Energy under contract DE-–AC05-–06OR23177. Further support was provided by the National Science Foundation, the Italian Istituto Nazionale di Fisica Nucleare, the Chilean Comisi\'on Nacional de Investigaci\'on Cient\'ifica y Tecnol\'ogica (CONICYT), the Chilean Agency of Research and Development (ANID), the French Centre National de la Recherche Scientifique, the French Commissariat \`{a} l'Energie Atomique, and the National Research Foundation of Korea.



\begin{thebibliography}{99}
\bibitem{Data-DOI}
    The data supporting this research is openly available from the University of York Pure repository with DOI: 10.15124/99b659e3-e612-45f1-83f9-9122c9aa711c.
\end{thebibliography}
\end{document}